\documentclass[sn-mathphys,Numbered]{sn-jnl}

\usepackage{graphicx}%
\usepackage{multirow}%
\usepackage{amsmath,amssymb,amsfonts}%
\usepackage{amsthm}%
\usepackage{mathrsfs}%
\usepackage[title]{appendix}%
\usepackage{xcolor}%
\usepackage{textcomp}%
\usepackage{manyfoot}%
\usepackage{booktabs}%
\usepackage{algorithm}%
\usepackage{algorithmicx}%
\usepackage{algpseudocode}%
\usepackage{listings}%
\usepackage{amsmath}
\usepackage{parskip}
\usepackage{setspace}
\usepackage{blindtext}
\usepackage{layout}
\usepackage{blindtext}
\usepackage{geometry}
\usepackage{amsmath} 
\usepackage{amsthm}
\usepackage[english]{babel}
\usepackage{amssymb,amsmath,amsthm, amsfonts}
\usepackage{csquotes}
\usepackage[hyphenbreaks]{breakurl}
\usepackage{hyperref}
\usepackage{xfrac}
\usepackage{longtable}
\usepackage{booktabs}
\usepackage{tabularx}
\usepackage{amsmath,amssymb,amsfonts,bm}
\usepackage{multirow}
\usepackage[font=small,labelfont=bf]{caption}
\usepackage{setspace}
\usepackage[usestackEOL]{stackengine}
\usepackage[T1]{fontenc}
\usepackage{graphicx}
\usepackage{adjustbox}
\usepackage{xcolor, colortbl}
\usepackage{todonotes}
\usepackage{gensymb}
\usepackage{adjustbox}
\usepackage{upgreek}
\usepackage{xcolor}
\usepackage{ulem}
\usepackage{tablefootnote}
\usepackage{lipsum}
\usepackage{textgreek}
\usepackage{amsmath}
\usepackage{amsfonts}
\usepackage{amssymb}
\usepackage{graphicx}
\usepackage{natmove}
\usepackage{wasysym}
\usepackage{lineno}
\usepackage{makecell}

\newgeometry{left=2cm}
\newgeometry{right=2.2cm}

\begin{document}
\title[Article Title]{GHz-rate optical phase shift in light matter interaction-engineered, silicon-ferroelectric nematic liquid crystals
}


\author*[1,2]{\fnm{Iman} \sur{Taghavi}}\email{staghavi3@ece.ubc.ca}
\author[1]{\fnm{Omid} \sur{Esmaeeli}}
\author[1]{\fnm{Sheri} \sur{Jahan Chowdhury}}
\author[3]{\fnm{Kashif} \sur{Masud Awan}}%
\author[1,2]{\fnm{Mustafa} \sur{Hammood}}
\author[1,3]{\fnm{Matthew} \sur{Mitchell}}
\author[1,3]{\fnm{Donald} \sur{Witt}}
\author[4]{\fnm{Cory} \sur{Pecinovsky}}
\author[4]{\fnm{Jason} \sur{Sickler}}%
\author[1,3,5]{\fnm{Jeff} \sur{Young}}
\author[1]{\fnm{Nicolas} \sur{A.F. Jaeger}}
\author*[1,2,6]{\fnm{Sudip} \sur{Shekhar}}\email{sudip@ece.ubc.ca}
\author*[1,2,3]{\fnm{and Lukas} \sur{Chrostowski}}\email{lukasc@ece.ubc.ca}

\affil[1]{\orgdiv{Department of Electrical and Computer Engineering}, \orgname{University of British Columbia}, \orgaddress{\street{2332 Main Mall}, \city{Vancouver}, \postcode{V6T 1Z4}, \state{B.C.}, \country{Canada}}}

\affil[2]{\orgdiv{Dream Photonics}, \orgaddress{\street{2366 Main Mall}, \city{Vancouver}, \state{B.C.}, \country{Canada}}}

\affil[3]{\orgdiv{Quantum Matter Institute}, \orgname{University of British Columbia}, \orgaddress{\street{2355 E Mall}, \city{Vancouver}, \postcode{V6T 1Z4}, \state{B.C.}, \country{Canada}}}

\affil[4]{\orgdiv{Polaris Electro-Optics}, \orgaddress{\street{3400 Industrial Ln, 3130 25th St}, \city{Broomfield}, \postcode{80020}, \state{CO}, \country{USA}}}

\affil[5]{\orgdiv{Department of Physics and Astronomy}, \orgname{University of British Columbia}, \orgaddress{\street{6224 Agricultural Road}, \city{Vancouver}, \postcode{V6T 1Z1}, \state{B.C.}, \country{Canada}}}

\affil[6]{\orgdiv{Dream Photonics Inc.}, \orgaddress{\city{Woodinville}, \state{WA}, \country{USA}}}


\abstract{\large\onehalfspacing Organic electro-optic materials have demonstrated promising performance in developing electro-optic phase shifters. Their integration with other silicon photonic processes, nanofabrication complexities, and durability remains to be developed. While the required poling step in electro-optic polymers limits their potential and utilization on a large scale, devices made of paraelectric nematic liquid crystals suffer from slow bandwidth. In ferroelectric nematic liquid crystals, we report an additional GHz-fast phase shift that ultimately allows for significant second-order nonlinear optical coefficients related to the Pockels effect. It avoids poling issues and can pave the way for hybrid silicon-organic systems with CMOS-foundry compatibility. We report DC and AC modulation efficiencies of $\approx$~0.25 V$\cdot$mm (from liquid crystal orientation) and $\approx$~25.7 V$\cdot$mm (from Pockels effect), respectively, an on-chip insertion loss of $\approx$~2.6 dB, and an electro-optic bandwidth of $f_\text{-6dB}$>4.18 GHz, employing improved light-matter interaction in a waveguide architecture that calls for only one lithography step.
}


\begin{spacing}{1.2}
\maketitle
\large
\section*{Introduction}
Silicon and silicon nitride photonics have enabled several recent technological advancements, and future roadmaps have been designed based on their capabilities and integration with established CMOS foundries [1]. The electro-optic phase shifter (EOPS) [2] is one of the most fundamental components upon which other essential building blocks in photonic integrated circuits (PIC) are based, including electro-optic (EO) switches [3], EO modulators [4], optical interconnects [5], and comb generators [6].
Traditional phase shift mechanisms often show poor overall performance, such as limited EO bandwidth (BW) and high static power consumption ($P_{\text{stat}}$) when leveraging the thermo-optic effect or free-carrier dispersion effect with travelling wave modulators. Free carrier dispersion also leads to doping-induced insertion loss (IL). Innovations in hybrid material platforms promise viable solutions to traditional Si photonics (SiP) shortcomings. However, they typically face significant challenges, such as several complex nanofabrication stages, scaling, and interaction with current CMOS foundries.
~Modulators based on indium tin oxide (ITO) have recently experienced notable improvements [7]. The additional operations necessary for ITO/insulator thin-film deposition, demanding quality control methods, and their low extinction ratio (ER) suggest further R\&D needs [8]. 
Carrier-free phase shift processes, such as the Pockels effect, are among the more appealing techniques, especially regarding EO bandwidth and energy consumption. These metrics and low IL can excel in LiNbO$_3$ on silicon (LNOI) [9-10]. Due to the low EO coefficient of LiNbO$_3$, however, phase shifters based on LNOI require either a large footprint or a high drive voltage ($V_{\uppi}$), where $V_{\uppi}$ is the voltage necessary to produce $\pi$ shift. A material system based on barium titanate (BTO) can simultaneously meet the requirements of low IL and high EO bandwidth [11-12].  
Despite having a significant EO coefficient, BTO is unable to fully utilize the improved light-matter interactions (LMI) provided by some waveguide topologies, such as slot waveguides, failing to give the record-low modulation efficiencies ($V_{\uppi}L$), where $L$ is the EOPS effective length [13]. BTO also involves meticulous fabrication steps, loses its high EO efficiency at cryogenic temperatures, and necessitates process controls for optimum domain development [14].\par 
Furthermore, the preceding approaches cannot meet the trade-offs between various device metrics of a particular modulator, such as footprint, EO and optical BW, IL, $V_{\uppi}$, and dynamic energy consumption ($E_\mathrm{dyn}$) [15-16].
Hybrid integration of Si with organic electro-optic (OEO) materials has been proposed to overcome these shortcomings. They have advantages over their inorganic equivalents due to a combination of their EO and physicochemical characteristics. The EO properties of OEO materials, including the Pockels factor ($n^{3}r_{33}$) and optical loss, have continuously progressed over the last two decades ($n^{3}r_{33}$ as large as 9000~pmV$^{-1}$ and absorption coefficients as low as $1.8\times10^{-5}$ have been reported [17], where $n$ is the OEO refractive index and $r_{33}$ represents the Pockels coefficient). Thanks to their liquid state, an OEO can infiltrate nanostructures during the coating step and benefit from the enhanced LMI. OEO's low dielectric dispersion [18] promises better modulators and the resultant high modulation efficiencies.
As a result, <1 V, sub-mm-long Mach-Zehnder modulators (MZM) with IL < 1 dB, EO bandwidth surpassing 100 GHz, and $E_\mathrm{dyn}<$100 $f$Jbit$^{-1}$ have been enabled [19]. Such an all-inclusive, miniaturized modulator can work in extreme temperatures (as high as 383 K (110$^\text{o}$C) [20] and as low as 4 K [21,22]), which encompass the crucial requirements for a state-of-the-art modulator.\par
While EO polymers have attracted increased attention mainly because of their remarkable EO characteristics, their application to photonic-integrated circuits (PIC) has challenges. Multiple process steps and often hazardous solvents are needed to prepare EO polymer thin films, necessitating equipment for precise mixing, coating, and drying. In addition, to align chromophores in an EO polymer material, an electro-thermal poling step is necessary. On a larger scale, dedicated DC lines may be required depending on the waveguide geometry and accompanying LMI. Despite considerable breakthroughs in producing thermally and temporally stable polymers, not all poled polymers are resilient to high temperatures ([20]), aging ([23]), and poling-induced optical loss ([24]). For practical applications, the devices may still require a certain level of hermetic sealing to become temperature- and humidity-resistant. EO polymers are also recognized for their lower resistance at higher temperatures, which coincides with detrimental leakage currents. It leads to decreased EO efficiency, which has a direct impact on device performance. Additional nanofabrication steps have been identified as advantageous to resolving this issue, such as exploiting a thin charge-barrier layer of alumina or titania created by atomic layer deposition [25,26]. Poling also limits high-temperature excursions typical of back-end-of-line processing since such excursions would destroy the poling-induced order. The problems above could increase the overall cost, add complexity, and jeopardize the integration of polymer-based hybrid platforms.\par
Another class of OEO material is PN-LC, which provides a variety of efficient but slow phase shift processes, such as the birefringence effect [27-30]. Ultra-high modulation efficiencies have been reported using strong-LMI implementations, including infiltrated slot waveguides with the material [2]. Unlike polymers, preparing PN-LC devices does not involve a solvent that must be evaporated during post-baking processes. It simplifies the spin coating process and may allow for alternative coating approaches. The material, for example, can be coated on photonic parts via microfluidic channels or fluid dispensing. A PN-LC-based platform is attractive for realizing EO switches, phase shifters, or shutters, but it is not desirable for high-speed applications.\par
Another state of matter, ferroelectric nematic liquid crystals (FN-LC), has recently been discovered [31-33].
In the landscape of organic EO materials, FN-LC is situated between poled polymers, organic nonlinear optic (NLO) crystals, and PN-LCs, bringing the most desirable aspects of each (see Table \ref{OEO}). Like poled polymers and organic NLO crystals, FN-LC is non-centrosymmetric and has nonzero second-order susceptibilities ($\upchi^{(2)}$). Like organic NLO crystals and PN-LCs, they organize spontaneously\textendash the ordered state is the thermodynamic minimum with significant order parameters. Similar to PN-LCs, the alignment can be controlled over large areas by applying alignment layers at the interface or small external fields. They can also be switched like PN-LCs, with a reorientation of the fast and slow axes in an external field. A unique feature of the FN-LC is that the molecular long axis (slow axis) is highly polarizable and coincident with the director and the polar axis (Fig. \ref{overview}(a) inset). This creates the condition for achieving very large spontaneous polarizations and large second-order nonlinearities, which have been observed by second harmonic generation measurements [34].
These nonlinearities have been dramatically enhanced by adding NLO dyes to the polar matrix of the FN-LC host material [35]. However, what remains to be demonstrated is a fast EO effect (i.e., Pockels effect) observed in other non-centrosymmetric organic materials. Various operational regimes for different variations of FN-LC were described in [36], where the authors explored other EO properties of the material but did not report the sub-nanosecond characteristic relaxation times or the Pockels effect we demonstrated here. The study found that $\tau{_\text{E}}< 100~\mu s$ may be attained with mild external electric fields ($\textbf{E}_\text{e}$). At larger $\textbf{E}_\text{e}$ values over 1 V$\mu m^{-1}$, their model predicted a smaller relaxation time inversely proportional to $\textbf{E}_\text{e}$, down to $\tau{_\text{E}}\approx 1.6~\mu s$.\par 
We report demonstrating a considerably faster EO effect in FN-LC, utilizing the nanoscale properties of our SOH structure. We specifically show that FN-LC can be integrated with similar ease as PN-LCs onto a SiP platform, that alignment of the polar axis can be easily achieved over the length of the device, and that a significant EO response can be observed. This paper proposes a hybrid material platform for PICs comprising numerous moderate-speed EOPS that prioritize large-scale scaling, integrability, and ease of fabrication. It includes non-modulation applications, such as switches [3,37], transducers [38], and weight banks [39] for neuromorphic PICs [40] and quantum computing [41].\par

\section*{Results}
\subsection*{LMI-engineered design}
Waveguides such as Si [42] and plasmonic [43] slotted, sub-wavelength grating (SWG) [44], photonic-crystal [45], and thinned [20], are traditionally employed as EOPS in modulators. However, they often add excessive loss [46], limit optical BW [46], and suffer from poor infiltration and poling [25,47]. Of particular interest for the current work is the extra fabrication stages slotted-based devices demand [38], some of which are not quite CMOS-compatible. For instance, opening oxide cladding over isolated Si islands of SWG or metallic slots of plasmonic waveguides may violate design rules in wafer nanofabrication. Also, exploiting the benefits of structures such as thinned waveguides necessitates taper converters from complete to partial etch Si [20]. Geometries required for efficient and low-loss mode converters, such as strip-to-slot, are not always accessible through nanofabrication runs or have integration challenges with the rest of the PIC. On the other hand, a non-slotted waveguide does not have these issues at the cost of offering a weak LMI [38], thus requiring a high voltage for poling and driving. The device size and EO bandwidth must then be compromised to keep the voltage within the ranges supplied by CMOS drivers.\par 

We propose a finger-loaded strip (FLS) waveguide consisting of two strip-to-finger mode converters to resolve this trade-off, as seen in Fig. \ref{overview}(a). In analogy to segmented slotted [48] and non-slotted [49] topologies, this structure eliminates the misalignment problem seen in strip-loaded slot waveguides by offering at least one fewer step for lithography and etching. The periodicity and duty cycle of the subwavelength structure of the FLS waveguide is restricted by fabrication capabilities yet selected to produce a better balance between electrical and optical loss penalties, another prevalent difficulty in slot waveguides. The waveguide core, sandwiched between fingers, is narrowed to $W_\text{c}=$~300 nm. The optical field confinement factor ($\beta$) of the localized light in the dual gaps between fingertips and the core is shown in Fig. \ref{Sims}(d), and the simulated values are provided in Methods.\par

Other than the so-called optical field confinement factor $\beta$, what determines the modulation sensitivity of a device based on the waveguide is the overlap integral ($\Gamma$) between optical ($\textbf{E}_\text{x}$) and electrical ($\textbf{E}_\text{e}$) modal fields. Considering the high $\Gamma$, the short effective distance between the two electrodes ($d_\text{eff}$) brought by the fingers adjacent to the core, the FLS waveguide offers competitive modulation efficiency-loss product (i.e., $V_\uppi L \upalpha$) compared to a non-slotted waveguide (see Methods). Besides, our simulations revealed that the 10~$\mu m$-long strip-to-finger mode converter does not substantially impact total EOPS loss (conversion efficiency~$\approx~99.86\%$). Thanks to its porous structure, FLS topology makes the most of the capillary force action to mitigate material infiltration, another ubiquitous issue in slot waveguides that frequently calls for oxide undercutting [50].\par 

\subsection*{Material handling}
The constructed devices must then be aligned with an electric field. Because of its pre-ordered molecular orientation and sensitivity to an external field, FN-LC was found to be fully aligned with $>5\sim10$ times smaller $\textbf{E}_\text{e}$ than required for poling EO polymers [51], in similar structures. The process was carried out at room temperature; no precise temperature monitor was necessary. These relaxed voltage and temperature conditions have additional advantages. First, an inert gas such as N$_2$ is not required around the chip during alignment to avoid oxidation-driven degradation. The FN-LC alignment process is relatively safer because there is no off-gassing. Within the scope of our characterization setup, the device's performance exhibited little temperature susceptibility thanks to the non-electro-thermal alignment technique. To be more precise, no performance aging was discovered for the FN-LC-coated devices over 150 hours of continuous operation in the range of 10$~\sim50^{\circ}$C under full optical and electrical load. EO polymers lose their poling efficiency beyond their glass transition temperature ($\approx80\sim100^{\circ}$C). To show that FN-LC is helpful at these temperatures, more accelerated aging tests must be done in the future. 
~The lower voltage required for alignment simplifies design concerns for the large-scale integration of FN-LC-enabled photonic devices in which the same RF lines can be utilized.\par

\subsection*{Characterization}
Monitoring optical loss and leakage current throughout the alignment process provides critical information. As seen in Fig. \ref{DC}(a)-(b), changes in optical loss followed by saturation indicate that domain formation has been completed. In some cases, the change was permanent, indicating that the optic axis continues to remain in a vertical posture. As shown in Fig. \ref{DC} (b), the fully-aligned FN-LC MZM showed $\approx~$2 to 3 dB less material absorption loss compared to the material with unaligned dipoles, which we attribute to the formation of a monodomain over the length of the device and reduced scattering loss (see Methods). The alignment-induced improvement in IL is opposite to that seen commonly in EO polymers, where the loss increases with voltage.\par

Following the alignment, we performed DC characterization of the optical transmission, as shown in Fig. \ref{DC}(c), achieving an optical ER of $\approx$~26~dB and $V_{\uppi,\text{DC}}\approx$~0.5~V. Various processes cause this significant shift, the most prominent of which is the birefringence effect. The optical transmission of the proposed MZM, shown in Fig. \ref{DC}(d), demonstrates an IL of $\approx2.1$ dB. Based on passive measurements of similar structures coated with a low-loss polymer featuring approximately the same index of refraction, we found $\approx1.6$ dB of the IL primarily attributed to the propagation loss through the non-optimized subwavelength structure of the deployed FLS waveguides and the rest $\approx0.5$~dB stemmed from material absorption at the wavelength of operation. Additionally, the birefringence effect contributes to an increase of around 1 dB, as indicated in Fig.\ref{DC}(d). Nonetheless, just a $V_\text{bias}=V_{\pi}/2$ is required during RF operation, implying an additional loss of $\approx(1/2)\times1$ dB, resulting in a total IL of $\approx 2.6$~dB. We also found that FN-LC has approximately the same absorption loss as an available EO polymer at the wavelength of operation. Aside from fabrication nonidealities, a physical path length difference of $\Delta L\approx 100~\mu m$ integrated into one of the arms has led to a free spectral range (FSR) of $6.6~nm$, as illustrated by Fig.~\ref{DC}(e). It agrees with the theoretical FSR of $\lambda^2/(n_\text{g} \Delta L)\approx6.3$ nm, where $n_\text{g}\approx 3.9$ is the group index of the FLS waveguide.\par

Pure phase efficiency is a seldom reported but very important figure of merit for EOPS for communications applications ($\eta_\text{pps}$). The metric is defined as $(\partial n / \partial v)~/(\partial \upkappa/\partial v$), where $n$ and $\upkappa$  are the real and imaginary parts of the complex refractive index ($n+i\upkappa$), respectively. $V_{\uppi}$ and loss can be correlated to the estimated values for $\partial n / \partial v$ and $\partial \upkappa/\partial v$. The optical ER, $V_{\uppi} L \upalpha$, $\eta_\text{pps}$ comprise a comprehensive assessment of EOPS appropriateness. Table \ref{FOMtable} summarizes the abovementioned figure of merits for an EOPS built on hybrid material platforms, including FN-LC and the conventional pn-junction technology. For our proposed FLS waveguide, significantly higher records up to $\eta_\text{pps}\approx$~88.9 have been achieved, which we credit largely to the material properties.\par

With a high-speed phase shift mechanism, FN-LC might demonstrate significant benefits over traditional PN-LC, in which the slower mechanisms mostly dominate the response. Various aspects of the frequency response characterization of the FN-LC-coated devices are shown in Fig. \ref{AC}, and the measurement details are covered in Methods. This step used SWG-based grating couplers (GC) for light injection into the devices [53-55]. The S-parameter collected by a Vector Network Analyzer (VNA) indicates that a -6dB EO bandwidth $f_\text{-6dB}>$~4.18 GHz. This achievement demonstrates a phase shift mechanism with a rise/fall time in the nanosecond range, while PN-LC is limited to functioning in the millisecond range [2]. Accordingly, we employed $S_{21}$ data and an analytical approach (see Methods) to evaluate the EO properties of the FN-LC material and the proposed EOPS. As a result, an EO coefficient of $r_{33}\approx$~24 pmV$^{-1}$ and an AC modulation efficiency of $V_{\uppi} L\approx$~25.7 (V$\cdot$mm) were estimated. The metrics are still significantly lower than the maximum record for EO polymers. This, being the first demonstration, promises further improvement in the future.\par

\begin{figure}
  \includegraphics[width=1\textwidth]{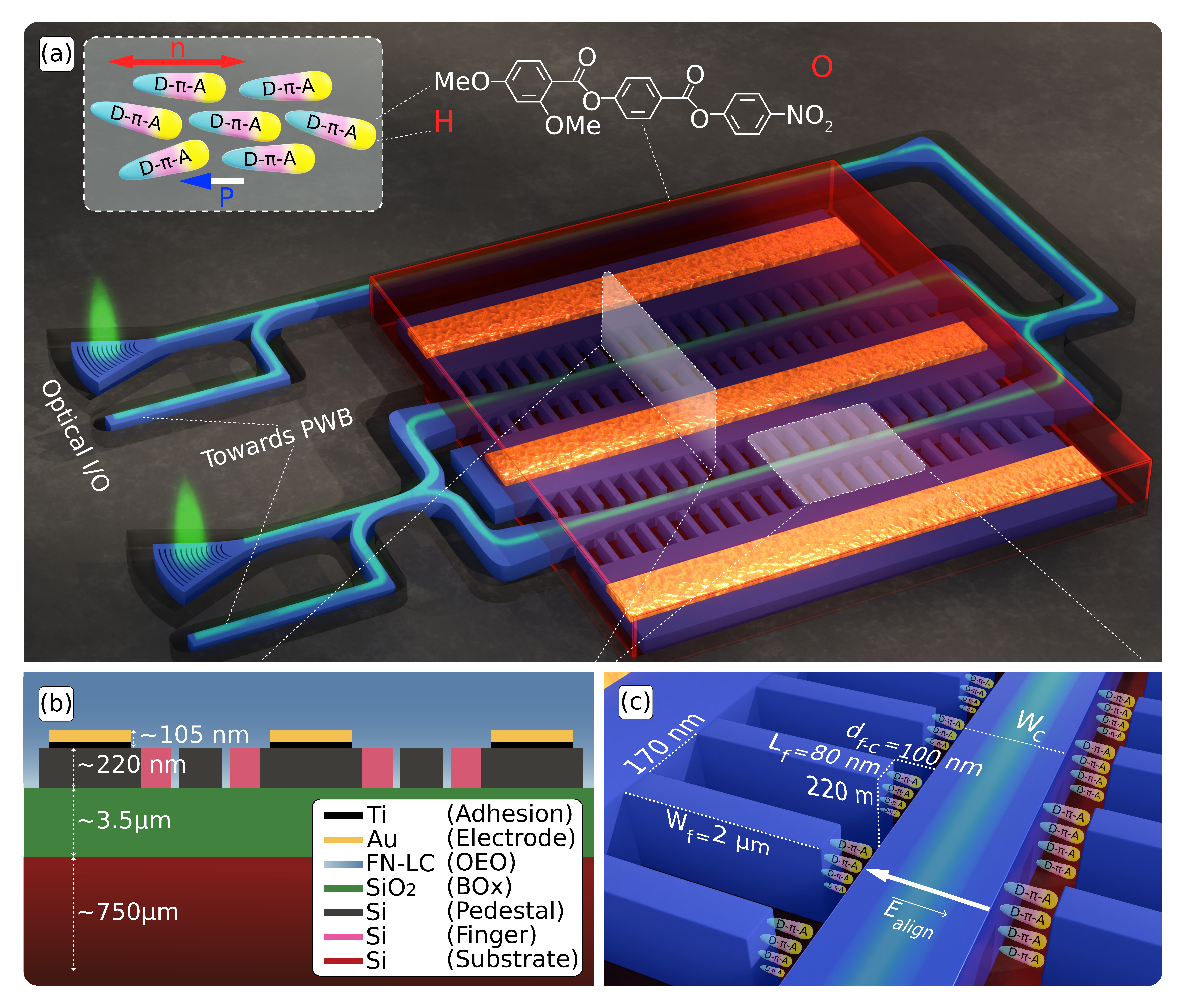}
  \caption{\textbf{3D representation of the modulator selectively covered by ferroelectric nematic liquid crystals (FN-LC).} \textbf{a} 80-nm-wide fingers are used to carry the RF signals and alignment field ($\textbf{E}_\text{align}$) to an area as close as $d_\text{f-c}=$~100 nm away from the $W_\text{c}=$~300 nm-wide core. This design optimizes the overlap integral ($\Gamma$) between the optical and electrical fields while minimizing optical loss. An external light source, either through a grating coupler or a photonic wirebond (PWB), feeds the optical signal. The waveguides are identified by a $2~\mu $m trench that surrounds them. A positive-tone ebeam resist patterns the device. \textbf{Inset:} An illustrative chemical structure and polar order of dipolar constituents of the FN-LC phase, where $n$ denotes the molecular director, $P$ is the polarization, and $D-\uppi-A$ is the donor-bridge-acceptor electronic structure. \textbf{b} The layer stack-up of the Mach-Zehnder Modulator (MZM) consists of two finger-loaded strip (FLS) waveguides in each arm. \textbf{c} A zoomed-in diagram of the FLS waveguide with infiltrated FN-LC ordered along $\textbf{E}_\text{align}$.}
  \label{overview}
\end{figure}

\begin{figure}
\includegraphics[width=1\textwidth]{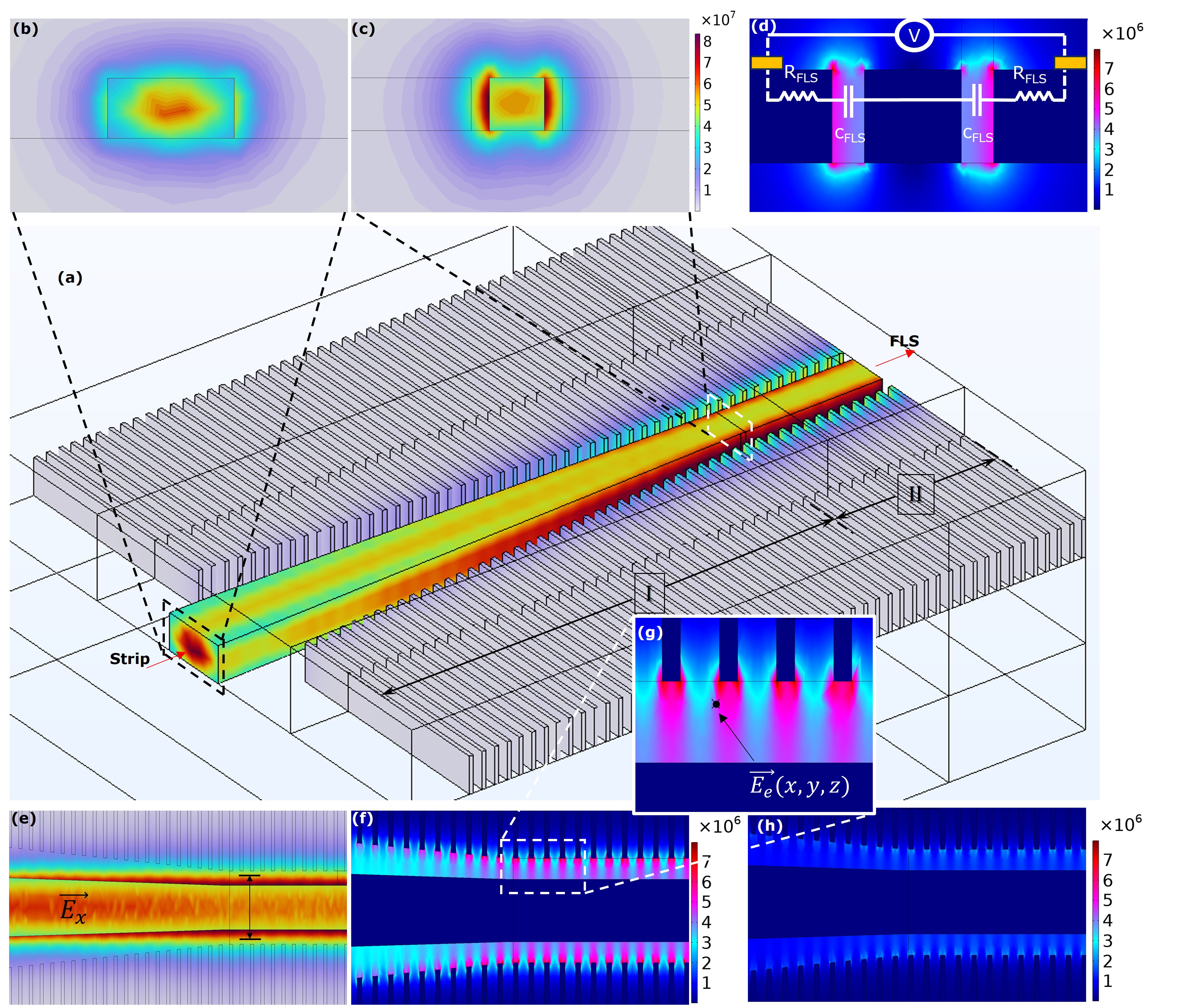}
\caption{\textbf{Waveguide design simulation results.} \textbf{a} The mode evolution in each arm of the Mach-Zehnder Modulator (MZM) demonstrated by the 10 $\mu m$-long adiabatic strip-to-finger mode converter (section I) coupled to a 3 $\mu m$-long portion of the 500 $\mu m$ finger-loaded strip (FLS) waveguide (section II). Transverse electric optical field ($\textbf{E}_\text{x}$) in x-slice sampled \textbf{b} before and \textbf{c} after the converter, which represents non-slotted and FLS waveguides, respectively. $\textbf{E}_\text{x}$ is decentralized from the narrowed core section ($W_\text{c}=$~300 nm) and pushed to the two gaps surrounding it, where the electrical field ($\textbf{E}_\text{e}$) and ferroelectric nematic liquid crystals (FN-LC) coexist. \textbf{d} Electric field ($\textbf{E}_\text{e}$) in x-slice sampled after the converter along with the equivalent RC circuit of the FLS architecture. $R_\text{FLS} (\approx 1.6~k\Omega)$ and $C_\text{FLS} (\approx 70.29~fF)$ represent the resistance of the finger and capacitance of the gap between the core and finger, respectively. Compared to an equivalent slot wavguide architecture (i.e., $W_\text{slot} = d_\text{f-c}$), the time constant is approximately halved ($\tau_\text{RC}=2R_\text{FLS}\times C_\text{FLS}/2$), neglecting the core resitance ($R_{c}$). Besides, the ratio of $\textbf{E}_\text{x}$ confined in the so-called gaps (i.e., $\beta$) and its overlap integral with $\textbf{E}_\text{e}$ are $\beta\approx15.4\%$ and $\Gamma\approx0.26$, respectively. \textbf{e} $\textbf{E}_\text{x}$ and \textbf{f} $\textbf{E}_\text{e}$ in the y-slice perspective (y~=~110~nm). Compared to a non-slotted waveguide, the light-matter interactions (LMI)-engineered FLS waveguide acts like a double-slot waveguide to offer $\approx$~21 times improvement in the $\textbf{E}_\text{e}$ as a result of decreased effective electrode distance ($d_{\text{eff}}$) as well as $\approx$~1.75 times higher $\Gamma$. $\textbf{E}_\text{x}/\textbf{E}_\text{x,max}$ drops to $1/e$ before it overlaps with the encompassing fingers. \textbf{g} A zoomed-in view of $\textbf{E}_\text{e}$ indicating that both $\textbf{E}_\text{e}$ and $\textbf{E}_\text{x}$ are evaluated at each coordinate (x,y,z) in the space, where light interacts with the FN-LC. The marked point proved to be a good reference when estimating $d_{\text{eff}}$ (see Methods). \textbf{h} $\textbf{E}_\text{e}$ evaluated at 4.18~GHz indicates the RF loss originated by low doping of the FLS waveguide ($\approx 5\times10^{14}~$cm$^{-3}$).}
\label{Sims}
\end{figure}

\begin{figure}
  \includegraphics[width=1\textwidth]{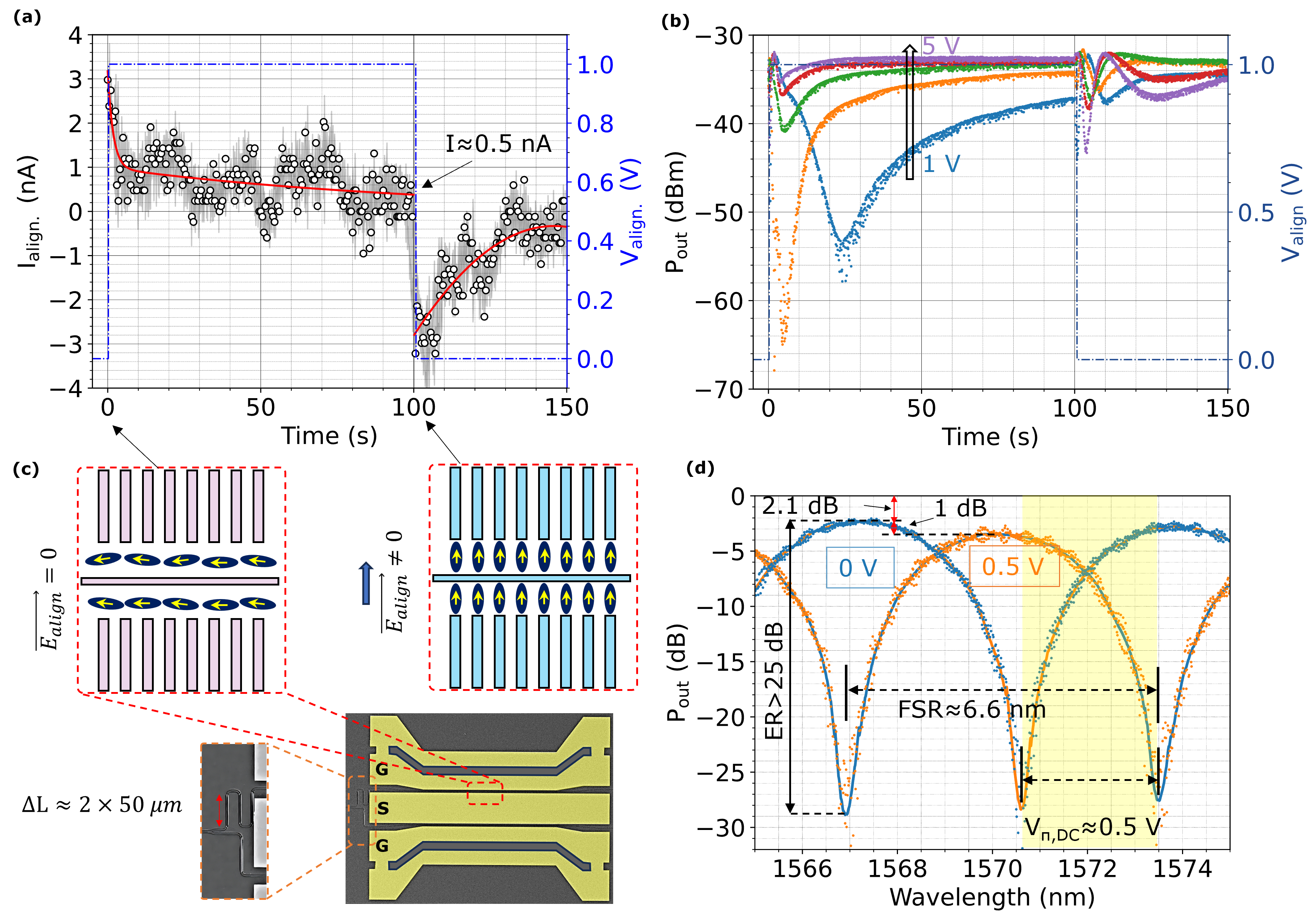}
  \caption{\textbf{ferroelectric nematic liquid crystals (FN-LC) alignment.} \textbf{a} Polarization-originated leakage current in response to an external electric field ($\textbf{E}_\text{align}~\propto~V_\text{align}/[2\times d_\text{f-c}]=2/(2\times0.1)~V/\mu m$). A static power of $P_\mathrm{stat}\approx0.5~nA\times2~V$ is then expected to maintain the complete alignment. \textbf{b} Optical output ($P_\text{out}$) versus increasing $\textbf{E}_\text{align}$. The polarity of $V_\text{align}$ is reverted at $t_\text{set}$. Despite the random changes in $P_\text{out}$ at the beginning, we figured a settlement time of $t_\text{set}\approx$100 s would be adequate to let dipole orientation relaxes and $P_\text{out}$ to saturate to some certain values associated with achieved molecular order. $P_\text{out}$ sampled at $t_\text{set}$ is $\approx~2\sim3~dB$ bigger once full alignment achieved. Therefore, insertion loss (IL) is reduced in FN-LC compared to its unaligned state (i.e., before applying any voltage). \textbf{c} A microscopic image of the MZM along with the dipolar order of FN-LC in a section of its finger-loaded strip (FLS) waveguide in the cases of $\textbf{E}_\text{align} \neq 0$ and $\textbf{E}_\text{align} = 0$ (other possibilities for dipolar order could exist, since we don't know what the ground state is with this topology). A physical path length difference of $\Delta L\approx100~\mu m$ is incorporated in the upper arm. \textbf{d} Optical output spectrum (normalized versus grating coupler loss) versus applied voltage exhibiting changes in the EOPS complex refractive index ($n+i\upkappa$ where $\triangle n \gg \triangle \upkappa$ for an ideal EOPS) to make a full $\pi$ shift enabled by the molecular orientation. The data is summarized in Table \ref{FOMtable}. A free spectral range (FSR) of $\approx6.6~nm$ is observed.} 
  \label{DC}
\end{figure}

\begin{figure}
  \includegraphics[width=1\textwidth]{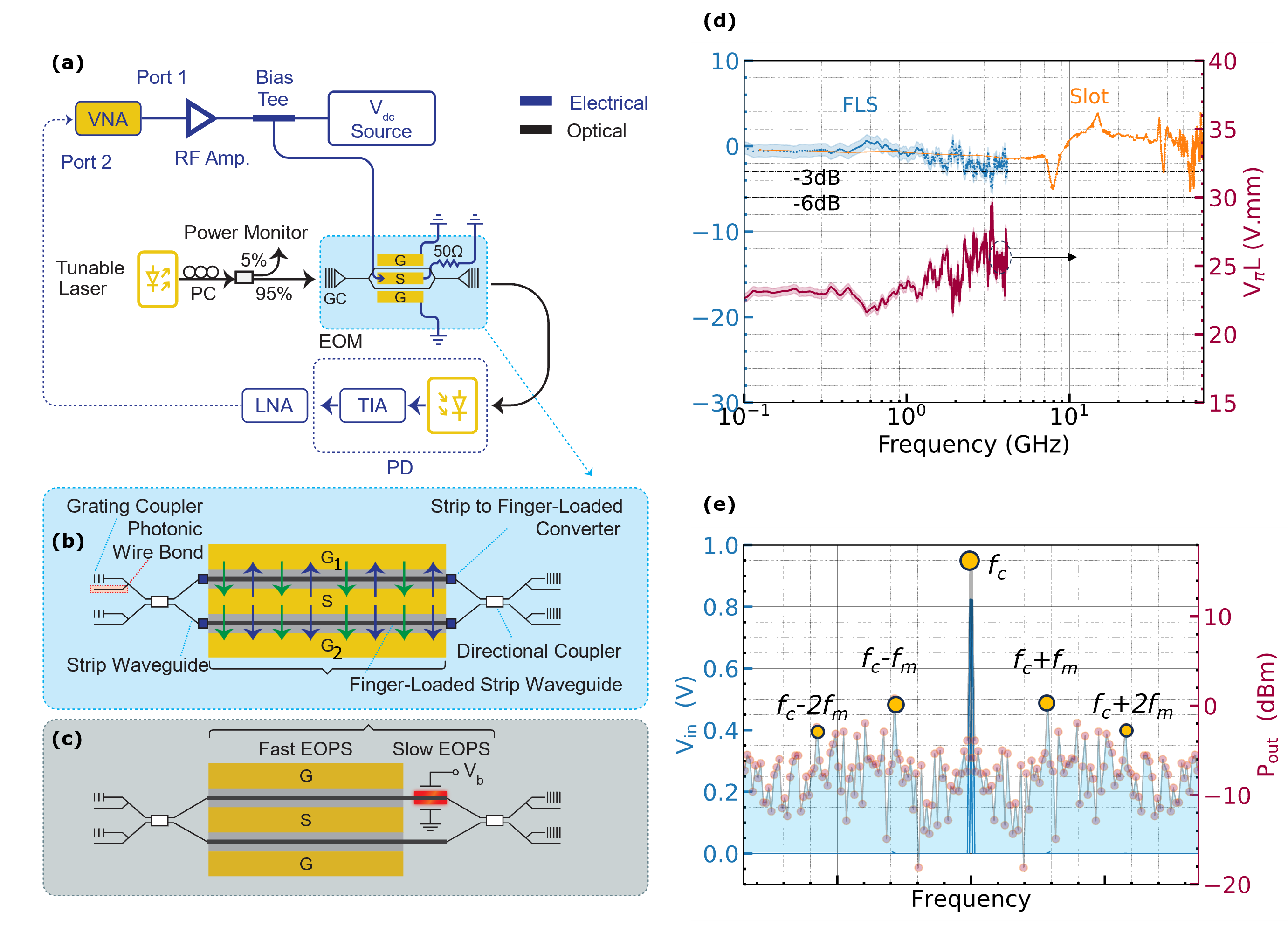}
  \caption{\textbf{Device characterizations.} \textbf{a} Experimental setup using a Vector Network Analyzer (VNA) and appropriate amplification to drive ferroelectric nematic liquid crystals (FN-LC) Mach-Zehnder Modulator (MZM) in push-pull configuration. The DC voltage ($V_\text{DC}$) applied to the bias Tee, mainly meant to bias the MZM at its quadrature point, can also be used to perform the alignment step (i.e., Fig. \ref{DC}\textbf{a}-\textbf{b}) before AC characterization test starts (see Methods). \textbf{b} MZM schematic showing two 500 $\mu $m-long EOPS and low-loss strip-to-finger-loaded mode converters equipped with GC and PWB optical I/O channels. The alignment has been performed unipolarly, as evidenced by the green arrows (i.e., the electrode labelled \textquotedblleft S\textquotedblright~ is grounded, while $V_\text{align}$ and $-V_\text{align}$ are the DC alignment voltages applied to \textquotedblleft G$_1$\textquotedblright~ and \textquotedblleft G$_2$\textquotedblright~ electrodes, respectively.) This configuration ensures that the FN-LC is aligned in the same direction in both arms. During regular RF operation, the signal is applied bipolarly to achieve push-pull performance, as the blue arrows show (i.e., it excites \textquotedblleft S\textquotedblright~, with \textquotedblleft G$_1$\textquotedblright~ and \textquotedblleft G$_2$\textquotedblright~ grounded). The proposed EOPS made of FN-LC incorporates both slow and fast phase shift mechanisms compared to the conventional approach, shown in \textbf{c}, in which a power-inefficient and slow effect (e.g. thermo-optic) biased by V$_{b}$ is required in conjunction with a fast EOPS for advanced modulation schemes. \textbf{d} The measured $|S_{21}|$ of the proposed undoped finger-loaded strip (FLS) device overlayed on that of a doped slot waveguide covered with the same FN-LC (data extracted from [67]). Post-processed AC values of $V_{\uppi}L$ based on extracted $r_{33}$ as a function of frequency are also shown. \textbf{e} Spectral analysis of the MZM optical output ($P_{out}$) driven by single tone sinusoidal wave ($V_{in}$ at $f_c = f_m \approx59$ MHz, where $f_c$ and $f_m$ are the carrier and modulation frequencies, respectively) verifying the existence of a fast phase shift enabled by the Pockels effect.}
  \label{AC}
\end{figure}

Considering the sub-mm phase shifters we used in each arm of the MZM, the mismatch between the optical and RF effective indices is not a limiting factor for the EO bandwidth [16]. The EO bandwidth is primarily limited by the weakly doped nature of fingers in the fabricated FLS waveguides. As shown by Fig. \ref{overview}, a -3dB EO bandwidth ($f_{-3dB}$) limited by the RC time constant ($\tau_{\text{RC}}$) of $\approx1.41$ GHz is anticipated, which agrees with Fig. \ref{AC}(d). As an immediate remedy, one could reduce finger lengths (i.e., $W_f$ in Fig. \ref{overview}) or place electrodes closer to the finger's end. As a secondary factor, the electrical wire bonds (EWB) that apply the RF signal could also have deteriorated the EO bandwidth. Unlike in slot waveguides, where some optical mode is confined in the two Si rails, less modal field propagates inside fingers in the FLS architecture. A single ion implantation step would consequently be sufficient to alleviate the poor $\tau_{\text{RC}}$ shown in this work while maintaining the RF-optical loss trade-off. Figure \ref{AC}(d) illustrates superimposed results from applying FN-LC to a properly doped slotted waveguide to validate this hypothesis (data extracted from [67]). Besides, as shown in Fig. \ref{Sims}(d), the FLS waveguide's twin slots reduce the overall $\tau_{\text{RC}}$. That means, less doping can alternatively achieve a desired $\tau_{\text{RC}}$, which instead would help improving the device's $V_{\pi}L\alpha$ (Supplementary Note 3).\par

We discovered that producing dependable EWBs with less complexity was relatively more straightforward than EO polymer-coated samples. The FN-LC phase shift mechanism appears to swing from a strong birefringence-driven effect at low frequencies to a weaker but faster Pockels effect at higher frequencies. Complex phase manipulation is usually required to develop advanced modulation systems, such as coherent transceivers [56]. Inefficient thermal heaters are frequently used to produce a slow phase shift (e.g., $\pi/2$), as illustrated in Fig. 4(c). Including slow and fast EOPS in a single material could be an appealing feature of the FN-LC-based platform.\par 

\subsection*{Large scale integration}

\begin{figure}
  \includegraphics[width=1\textwidth]{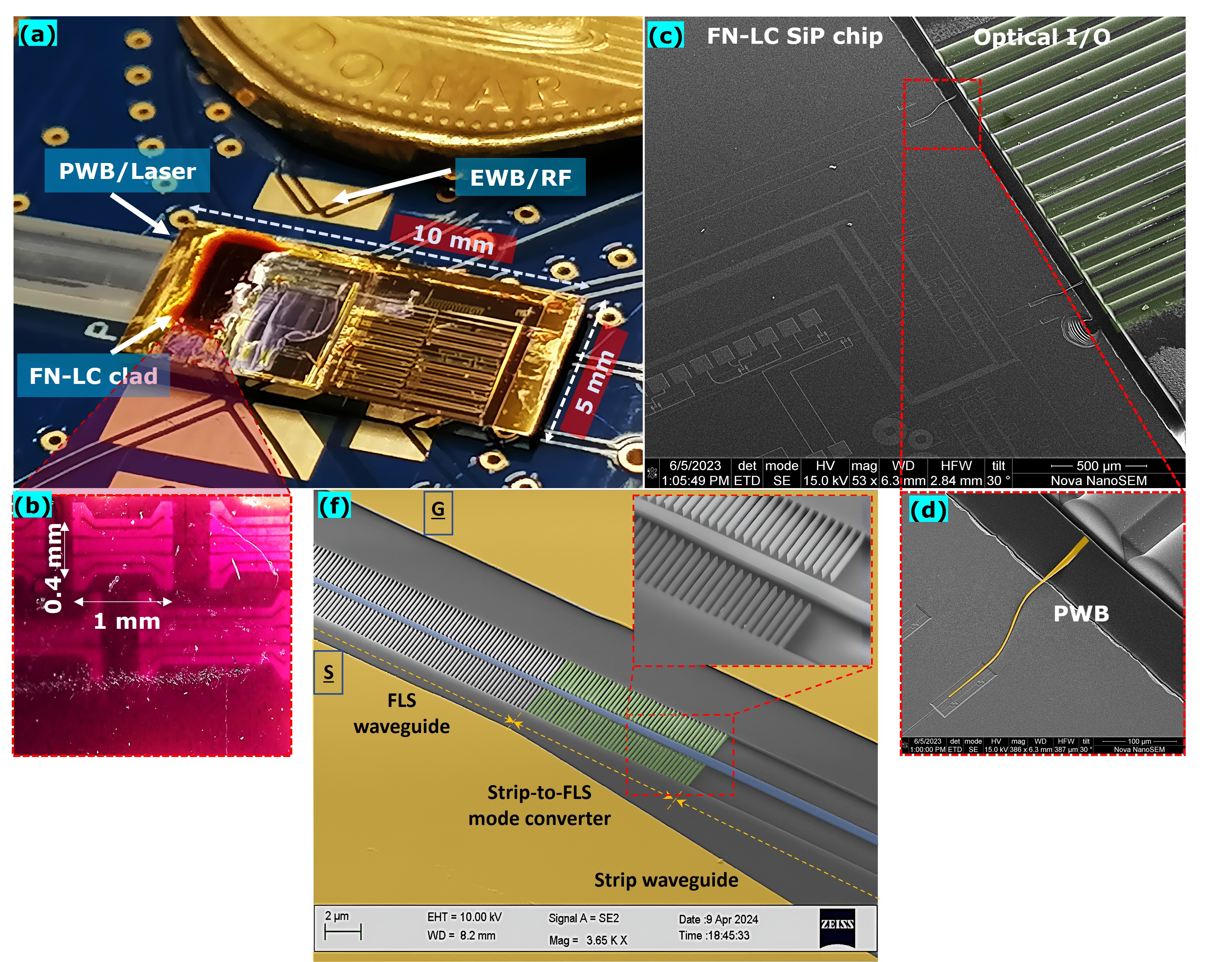}
  \caption{\textbf{Ferroelectric nematic liquid crystals (FN-LC)-enabled silicon photonic chip.} \textbf{a} A close-up picture of the chiplet incorporating 108 FN-LC Mach-Zehnder Modulator (MZM)s, the majority of which occupies a footprint of 0.4 mm$^{2}$. EWBs are used to carry RF signals at the chip level. FN-LC has been selectively applied to a cluster of MZMs on the left for demonstration purposes. Top-down GC and the PWB handle optical I/O for device and chip-level tests, respectively. \textbf{b} A microscopic zoomed-in view of the FN-LC-enabled modulators of various EOPS lengths. \textbf{c} A scanning electron microscope (SEM) picture shows PWBs as extra optical I/O paths and their adiabatic surface tapers for low-loss power coupling. A 3D laser printer tool (Vanguard Automation - Sonata 1000) uses the two-photon polymerization effect to create free-form optical waveguides. The FN-LC needs to cover the EOPS sections on the SiP chip, leaving adequate space for the polymer required for PWB construction. \textbf{d} A magnified, false-coloured SEM view of the PWB waveguide created between fibres and the on-chip surface taper. \textbf{f} A false-coloured SEM image of the light-matter interaction-engineered finger-loaded strip waveguide with a zoomed-in view of its mode converter.}
  \label{packaging}
\end{figure}

We evaluated the potential large-scale capability of our EOPS structures by creating a SiP chip with 108 MZMs on a 0.5~mm$^2$ area, as depicted in Fig. \ref{packaging}. A subset of devices was photonically wire-bonded (PWB) to an external laser source via a 16-channel fiber array (FA) block for demonstration purposes. The PWB routes were established using adiabatic surface tapers at the chip edge, followed by selective patterning of FN-LC on active components in a post-processing step. Because of its high viscosity, FN-LC can be applied locally, allowing for more controlled integration with the rest of the chip's components. Thanks to the reduced alignment voltage enabled by utilizing FLS waveguides in our Si-FN-LC design, the step can be paired with a $V_\text{DC}$, which is used to bias modulators at their quadrature point. Common EWB and a printed circuit board (PCB) were used to conduct alignment and drive the modulators. The demonstrated FN-LC SiP chip can be viewed as a prototype towards large-scale integration of organic material-enabled PICs, which we credit partly to overcoming the poling difficulties by utilizing the poling-free materials described here. A hypothetical EO polymeric version of this chip necessitates on-chip routing considerations and a separate PCB with potentially higher-voltage handling capabilities for poling.\par

The dual-phase shift of FN-LC provides a significant benefit for large-scale integration of dense EOPSs, with an estimated $P_\text{stat}\leq1~n$W required to maintain perfect alignment and drive the MZM at the quadrature point. At the frequencies of interest in this work, the proposed MZM's comparatively short arm length of $0.5$ mm eliminates the necessity for doubly-terminated travelling wave electrodes, another source of static power consumption in MZMs. As a result, the MZM can be driven like a lumped element device with no need for DC routes intended to supply doped heaters. It may also be helpful when numerous devices must coexist in a limited footprint, posing the risk of thermal crosstalk during individual device operations for conventional MZMs. In the case of EO polymers, a scalable method for introducing local heat that does not de-pole neighbouring devices has yet to be established. The poling-free nature of FN-LC mitigates this issue by eliminating the necessity for a locally raised energy level. Each device can be activated using dedicated RF lines routed for normal functioning, which is of special interest for addressing PICs in many applications (e.g., cryogenic), including those with a limited number of total RF and DC I/O lines.\par

Towards a truly integrated SiP-FN-LC chip on a large scale, one would require to show the cooperation of several EOPs in relevant scenarios such as the one reported in [57], which require a comprehensive electrical packaging scheme and will be covered in future work. Previously, we have also successfully integrated other photonic components, such as distributed feedback laser [58] and semiconductor laser amplifier [59], using a similar photonic packaging technique shown in this work. Furthermore, we anticipate some sealing would be beneficial to accomplishing a properly packaged FN-LC SiP chip despite our early observations of the insensitivity of FN-LC to agents such as humidity and oxygen.\par

\section*{Discussion}
We demonstrated GHz-fast EO phase shift and modulation in FN-LC. In contrast to EO polymers, we established that the material does not require any electro-thermal poling processes for alignment. Indeed, the material can be oriented during regular modulator RF operation. An FLS waveguide infiltrated with this material is proposed to reduce the number of lithography steps compared to slot waveguides while benefiting from improved LMI compared to non-slot ones. Using FLS phase shifters as small as 500 $\mu m$ in a push-pull MZM, we achieved a DC modulation efficiency of $V_{\uppi}L\approx$~0.25~V$\cdot$mm, an optical ER of $\approx$~26~dB, and an IL of $\approx$~2.6~dB. The propagation loss of $\approx~4.2~ \text{dBmm}^{-1}$ is similar to some slot waveguides [60-61]. We discovered that the material absorption contributes $\approx$~0.5~dB of the total IL and is very similar to the case of an EO polymer. The majority of the loss is attributed to the non-optimized FLS waveguide. Reducing fabrication imperfections, including less core sidewall roughness and better infiltration of FN-LC into the two spaces between the core and fingers, can further improve IL.\par

A significant phase shift efficiency in the order of $\partial n/\partial \upkappa \approx$ 88.9 was estimated, which we attribute to the material and the FLS waveguide employed in this work. Most importantly, we discovered a high-speed phase shift mechanism $f_\text{-6dB}$ of at least 4.18 GHz that outperforms the more substantial but slower birefringence effect in classic paraelectric-nematic liquid crystals. The EO bandwidth of our FN-LC-covered modulator is limited by the experiment setup and device fabrication, not the suggested architecture or the material, both of which are easy to fix but are out of the scope of this work. We established the existence of a Pockels-based linear electro-optic effect in FN-LC materials.\par

Also, an analytical approach was introduced to estimate the Pockels coefficient of the material, where an AC modulation efficiency of $V_{\uppi}L\approx$~25.7~V$\cdot$mm associated with a modest $r_\text{33}\approx$~24 pmV$^{-1}$ at $f = 4.18$ MHz. The difference between $V_{\uppi}L$ at DC and AC originates from the phase shift mechanism, which is dominant at each regime. At DC, the molecular orientation causes a significant modulation sensitivity (i.e., $\partial n/\partial V$). However, the faster Pockels effect dominates at higher frequencies, influenced by the modest $r_{33}$ reported above. Doping silicon waveguides can also lower $R_\text{FLS}$, resulting in a higher voltage drop ratio over the LMI region (i.e., $1/|1+2\textit{j}\pi\textit{f}\times R_\text{FLS}C_\text{FLS}|$, where \textit{f} is the operating frequency, $R_\text{FLS} $ and $C_\text{FLS}$ represent the resistance of the finger and capacitance of the gap between the core and finger, respectively). The effect is less significant unless at low frequencies, at which $f/f_{-3dB}\ll1$, where $f_{-3dB}=1/\big(2\pi R_\text{FLS}C_\text{FLS}\big)$.

As a sufficiently reliable and stable material for practical applications that can scale, the comparatively low modulation efficiency at higher speeds for this first-ever prototype is still a breakthrough and a harbinger of further improvements. Other waveguide embodiments featuring stronger LMI can be coated with FN-LC to achieve high AC modulation efficiencies at the cost of more involved fabrication processes.
The coexistence of a high-speed and effective but slow phase shift mechanism in FN-LC could pave the way for poling-free silicon organic PICs that have similar device metrics to those of EO polymers but require fewer strict steps to be made. The non-thermal nature of the alignment step (i.e., poling-free) allows for the selective alignment of phase shifters without affecting adjacent devices, which is one of the main benefits of this yet-to-be-thoroughly validated result. Future research will explore the performance of our proposed poling-free Si-organic modulators based on FN-LC for high-speed data transmission, switching, or transduction at cryogenic temperatures. The material will likely lose its beneficial DC phase shift as temperature decreases. Besides, alignment should be maintained during the cool-down process. Nonetheless, one may leverage the potentially better material's elasticity than EO polymers. It could be advantageous to relax the tension caused by different thermal expansions in such a SiP chip without requiring surface passivation or material alterations. Future research will examine the material's EO properties (i.e., $n$, $\upkappa$ and $r_{33}$) at cryogenic temperatures.

\section*{Methods}

\textbf{Device fabrication:} The modulators were patterned on a standard 220 nm silicon on an insulator wafer with a buried oxide layer of 3.5 $\mu m$. The pattern was first corrected for proximity effects using GenIsys Beamer software, applying both short- and long-range proximity effect corrections. The silicon-on-insulator (SOI) sample was cleaned using a standard solvent cleaning process with acetone and isopropyl alcohol (IPA), followed by nitrogen drying and a 30-second oxygen plasma treatment (20 sccm O2 flow). The sample was then baked at 180$^{\circ}$C for dehydration. Zep 520A-7, a positive-tone electron beam lithography (EBL) resist, was spin-coated at 2000 rpm with 1000 rpm/s for 35 seconds to achieve a resist thickness of approximately 260 nm. The sample was subsequently baked at 180$^{\circ}$C to remove any residual solvent. [54-55].\par

Resist exposure was carried out using a 100 keV EBL system (JEOL JBX-8100FS, JEOL) with a dosage of 213 $\mu \text{C}/cm^2$, a beam current of 1 nA, and a shot pitch of 4 nm. Development was performed in N-Amyl Acetate (ZED-N50, Zeon Specialty Materials Inc.) for 1 minute. The pattern was then transferred into the silicon device layer using an inductively coupled plasma reactive ion etching (ICP-RIE) process (Oxford COBRA). Etching parameters included a chamber pressure of 10 mTorr, a temperature of 15$^{\circ}$C, an ICP power of 600 W, a bias power of 30 W, and a bias voltage of 220 V. The gas mixture consisted of 30 sccm of C4F8 and 25 sccm of SF6, with a total etching duration of 3.5 minutes. Residual Zep resist was removed by exposing the sample to deep ultraviolet light for 15 minutes, followed by immersion in a heated PG remover bath for 15 minutes. The sample was then rinsed in IPA and dried with nitrogen.\par 

For metallization, the sample was treated again with a 30-second oxygen plasma (20 sccm O$_2$ flow) and baked at 110$^{\circ}$C for dehydration. AZ514E photoresist was used in image reversal mode for the lift-off process. Spin coating was performed at 4000 rpm with an acceleration of 1000 rpm/s for 40 seconds. After a post-bake at 110$^{\circ}$C for 1 minute, the sample was exposed using a maskless aligner (MLA-150, Heidelberg Instruments) with a 365 nm wavelength and an exposure dose of 40 mJ/cm$^2$. The sample underwent a second bake at 90$^{\circ}$C for 1 minute, followed by a flood exposure. Development was carried out using a dilute TMAH solution (AZ MIF 300 Developer, MicroChemicals) for 1 minute, followed by a 1-minute rinse in deionized water. Metal electrodes (100 nm gold atop 5 nm titanium as an adhesion layer) were deposited 4.85 $\mu $m distant from the core waveguide. A passive optical measurement was performed after completing the nanofabrication stages of photonic-only structures and after metallization. No evidence of EO activity was observed before coating FN-LC, which ensures no detectable contribution from free carrier or plasma dispersion effects.\par

After metallization, the chip is cleaned gently using acetone/IPA and an additional step of DI water for removing any water-soluble residues, dried with N$_2$ gun, and surface-passivated by a step of oxygen plasma treatment for $\approx$~30~sec just before coating. For the OEO cladding, we used PM-158 (a proprietary FN-LC material optimized for enhanced $\chi^{(2)}$ obtained from Polaris Electro-Optics, Inc.). According to our pre-coating viscosity experiments, the cleaned chip benefitted from being warmed up (70$\sim80^{\circ}$C) to reduce the viscosity of FN-LC during application to the chip. A glass capillary tube was used to apply enough FN-LC to the desired areas. Finally, the chip is cooled to 25$^{\circ}$C before characterization.\par

\noindent

\noindent
\textbf{AC characterization:}
The same SMU is used as a DC source with a set voltage precision of 0.02\% + 24 mV, which is 20 times greater than $V_{\pi, DC}=0.5~V$, thus allowing precise adjustment of MZM bias despite the efficient DC detuning mechanism in FN-LC. 
An 18 GHz vector network analyzer (Keysight FieldFox VNA) was added for AC characterization, with port 1 amplified using a microwave amplifier (SHF-s807). It is mixed with a proper amount of DC voltage for optimum MZM performance and simultaneously for FN-LC alignment. A high-speed photodetector (Thorlabs RXM40AF) was employed at the output, along with low-noise amplifiers (picosecond 5828). The VNA data was utilized to evaluate the device S-parameters. The devices were aged at a regulated temperature (25$^{\circ}$C) for $\approx$ 150 hours with no humidity control or hermetic sealing applied, with light ($P_\text{opt,in}$ = 8 dBm) and an RF electrical signal ($v_{\text{RF}}$ = 1.59 V) applied. In Fig.~\ref{AC}(e), an arbitrary function generator drives the amplifier, and the data captured by a digital storage oscilloscope (SDS6204A by SIGLENT) performs fast Fourier transformation.

The material EO coefficient of FN-LC (i.e., $r_{33}$) was post-processed using the analytical formula [52] (Supplementary Note 1)

\begin{equation}
        S_{21} = 20\times log\Biggl[\upalpha_\text{opt}\times\frac{\partial p_\text{mod}}{\partial n}\frac{\partial n}{\partial v}\frac{\partial v_\text{det}}{\partial(p_\text{det})}\times~G_\text{RF}\Biggr]
\end{equation}

\vspace{10pt}
\noindent where $G_\text{RF}$ is the calibrated gain of the modulator link, $\upalpha_\text{opt}$ is the total optical loss, $p_\text{mod}$ is the optical output of the MZM, $p_\text{det}$ ($v_\text{det}$) is the optical (electrical) input (output) of the detector, $n$ is the effective index of the FLS waveguide covered with FN-LC, and $v$ is the MZM drive voltage, respectively. Here, $\partial v_\text{det}/\partial p_\text{det}$ and $\partial n/\partial v$ correspond to the detector and the modulation sensitivities. The latter is related to $V_{\uppi}$ and, within good accuracy, to $r_{33}$ by:

\begin{equation}
        V_{\uppi}=\frac{\uplambda}{2L}\Biggl[\frac{\partial n}{\partial v}\Biggr]^{-1},
        \label{Vpi}
\end{equation}
\begin{equation}
        r_{33}=\frac{2d_{\text{eff}}}{n^3\Gamma}\Biggl[\frac{\partial n}{\partial v}\Biggr]
        \label{r33}
\end{equation}

\vspace{10pt}
\noindent In Equation \ref{r33}, $d_{\text{eff}}$ denotes the effective distance between the tip of two encountering fingers, which is inversely proportional to the average electric field ($\textbf{E}_\text{e,avg}$) over the LMI volume, i.e., where the optical ($\textbf{E}_\text{x}$) and electrical ($\textbf{E}_\text{e}$) modal fields coexist with the FN-LC.
Also, $\Gamma$ is the overlap integral between $\textbf{E}_\text{x}$ and $\textbf{E}_\text{e}$ [64] evaluated by:

\begin{equation}
        \Gamma=\frac1{Z_0}\iiint_V n_\text{eff}\left(\textbf{E}_\text{e}\right)\left\vert {\widehat E}_x\right\vert^{2}/\iint_A L\times \text{Re}\left(\textbf{E}\times\textbf{H}^\ast\right)
        \label{Gamma}
\end{equation}

\addtocounter{equation}{-1}\ignorespaces 
\vspace{10pt}
\noindent In Equation~\ref{Gamma}, \textit{n}\ensuremath{_\text{eff}} is the field-dependent effective mode index of FLS waveguide, \textit{A} and \textit{V} are the cross-section and volume of the LMI region, \textit{Z}\ensuremath{_{0}} is the free space impedance, and $\widehat H$ is the magnetic field of the optical mode, respectively. As a first approximation, one can assume $\Gamma\approx\beta$ [65,66]. To obtain a more precise value for $\Gamma$, an electrical field simulator calculates the local values of $\textbf{E}_\text{e}$, the results of which are called by an optical mode solver (COMSOL Multiphysics) to incorporate the field dependency of $n_\text{eff}$. We assumed FN-LC has similar permittivities at the optical and microwave frequencies in analogy to EO polymers [70]. Both electrical and optical modal field solvers are in 3D to consider the effect of the periodic structure of the FLS waveguide along the light propagation. The semi-double slot waveguide structure of the FLS waveguide can offer $\beta\approx0.15$ compared to $\beta\approx0.19$ of a slot waveguide ($W_{\text{slot}}$=100 nm) (Supplementary Note 2).\par 

However, the characteristic figure of merit of a phase shifter ($V_\uppi L\upalpha$) is $\propto\Gamma\times\upalpha/d_{\text{eff}}$. We have also compared a hypothetical non-slotted waveguide with a 300-nm core width surrounded by 2.1 $\mu m$ trenches. Our simulations revealed $\Gamma\approx$~0.26, 0.27, 0.15 for the proposed FLS waveguide, slotted and non-slotted structures, respectively. If doped properly, one can assume $d_{\text{eff}}\approx2\times d_{\text{f-c}}$ = 200 nm, i.e., 21 times smaller than that of a non-slotted structure. Our 3D point-by-point evaluations of $\textbf{E}_\text{e}$ agreed well with this approximation. On the other hand, we have recorded a total propagation loss of $\approx4.2$~dB/mm for the FLS waveguide, versus $\approx1$~dB/mm for the above-described, hypothetical non-slotted structures within the best etching and coating quality we could achieve. Using these three ingredients, one can conclude an improvement of $\approx~8.7$ times compared to a non-finger-loaded waveguide. The metrics are estimated to be still at least twice higher for the aforementioned slot waveguide thanks to smaller $d_{\text{eff}}\approx~W_{\text{slot}}$, at the cost of the described challenges.

\section*{Data availability}

The supporting data for the conclusions of this investigation are available upon reasonable request.

\section*{Code availability}

The algorithms used for this study are standard and are outlined in “Methods.” The corresponding authors can provide code scripts upon reasonable request.

\end{spacing}

\section*{References}
\normalem
[1] Shekhar, S., Bogaerts, W., Chrostowski, L., Bowers, J.E., Hochberg, M., Soref, R., Shastri, B.J. Roadmapping the next generation of silicon photonics. \textit{Nature Communication} \textbf{15}(751), 1–15 (2024).

[2] Xing, Y., Ako, T., George, J.P., Korn, D., Yu, H., Verheyen, P., Pantouvaki, M., Lepage, G., Absil, P., Ruocco, A., et al. Digitally controlled phase shifter using an soi slot waveguide with liquid crystal infiltration. \textit{IEEE Photonics Technology Letters} \textbf{27}(12), 1269–1272 (2015).

[3] Enami, Y., Luo, J., Jen, A.K. Short hybrid polymer/sol-gel silica waveguide switches with high in-device electro-optic coefficient based on photostable chromophore. \textit{Aip Advances} \textbf{1}(4), 042137 (2011).

[4] Tahersima, M.H., Ma, Z., Gui, Y., Sun, S., Wang, H., Amin, R., Dalir, H., Chen, R., Miscuglio, M., Sorger, V.J. Coupling-enhanced dual ito layer electro-absorption modulator in silicon photonics. \textit{Nanophotonics} \textbf{8}(9), 1559–1566 (2019).

[5] Sinatkas, G., Christopoulos, T., Tsilipakos, O., Kriezis, E.E. Electro-optic modulation in integrated photonics. \textit{Journal of Applied Physics} \textbf{130}(1), 010901 (2021).

[6] Gaeta, A.L., Lipson, M., Kippenberg, T.J. Photonic-chip-based frequency combs. \textit{Nature Photonics} \textbf{13}(3), 158–169 (2019).

[7] Amin, R., Maiti, R., Gui, Y., Suer, C., Miscuglio, M., Heidari, E., Chen, R.T., Dalir, H., Sorger, V.J.: Sub-wavelength ghz-fast broadband ito mach–zehnder modulator on silicon photonics. \textit{Optica} \textbf{7}(4), 333–335 (2020).

[8] Amin, R., Maiti, R., Carfano, C., Ma, Z., Tahersima, M.H., Lilach, Y., Ratnayake, D., Dalir, H., Sorger, V.J. 0.52 v mm ito-based mach-zehnder modulator in silicon photonics. \textit{Apl Photonics} \textbf{3}(12) (2018).

[7] Gui, Y., Nouri, B.M., Miscuglio, M., Amin, R., Wang, H., Khurgin, J.B., Dalir, H., Sorger, V.J. 100 ghz micrometer-compact broadband monolithic ito mach–zehnder interferometer modulator enabling 3500 times higher packing density. \textit{Nanophotonics} \textbf{11}(17), 4001–4009 (2022).

[8] Amin, R., George, J., Sun, S., Lima, T., Tait, A.N., Khurgin, J., Miscuglio, M., Shastri, B.J., Prucnal, P.R., El-Ghazawi, T., et al. Ito-based electro-absorption modulator for photonic neural activation function. \textit{APL Materials} \textbf{7}(8), 081112 (2019).

[9] He, M., Xu, M., Ren, Y., Jian, J., Ruan, Z., Xu, Y., Gao, S., Sun, S., Wen, X., Zhou, L., et al. High-performance hybrid silicon and lithium niobate mach–zehnder modulators for 100 gbit s-1 and beyond. \textit{Nature Photonics} \textbf{13}(5), 359–364 (2019).

[10] Wang, C., Zhang, M., Chen, X., Bertrand, M., Shams-Ansari, A., Chandrasekhar, S., Winzer, P., Lončar, M. Integrated lithium niobate electro-optic modulators operating at cmos-compatible voltages. \textit{Nature} \textbf{562}(7725), 101–104 (2018).

[11] Abel, S., Stöferle, T., Marchiori, C., Rossel, C., Rossell, M.D., Erni, R., Caimi, D., Sousa, M., Chelnokov, A., Offrein, B.J., et al. A strong electro-optically active lead-free ferroelectric integrated on silicon. \textit{Nature communications} \textbf{4}(1), 1671 (2013).

[12] Eltes, F., Mai, C., Caimi, D., Kroh, M., Popoff, Y., Winzer, G., Petousi, D., Lischke, S., Ortmann, J.E., Czornomaz, L., et al.. A batio 3-based electro-optic pockels modulator monolithically integrated on an advanced silicon photonics platform. \textit{Journal of Lightwave Technology} \textbf{37}(5), 1456–1462 (2019).

[13] Abel, S., Eltes, F., Ortmann, J.E., Messner, A., Castera, P., Wagner, T., Urbonas, D., Rosa, A., Gutierrez, A.M., Tulli, D., et al. Large pockels effect in micro-and nanostructured barium titanate integrated on silicon. \textit{Nature materials} \textbf{18}(1), 42–47 (2019).

[14] Eltes, F., Villarreal-Garcia, G.E., Caimi, D., Siegwart, H., Gentile, A.A., Hart, A., Stark, P., Marshall, G.D., Thompson, M.G., Barreto, J., et al. An integrated optical modulator operating at cryogenic temperatures. \textit{Nature Materials} \textbf{19}(11), 1164–1168 (2020).

[15] Miscuglio, M., Adam, G.C., Kuzum, D., Sorger, V.J. Roadmap on material-function mapping for
photonic-electronic hybrid neural networks. \textit{APL Materials} \textbf{7}(10), 100903 (2019).

[16] Taghavi, I., Moridsadat, M., Tofini, A., Raza, S., Jaeger, N.A., Chrostowski, L., Shastri, B.J., Shekhar,
S. Polymer modulators in silicon photonics: review and projections. \textit{Nanophotonics} \textbf{11}(17), 3855–3871347 (2022).

[17] Xu, H., Elder, D.L., Johnson, L.E., Heni, W., Coene, Y., De Leo, E., Destraz, M., Meier, N., Vander Ghinst, W., Hammond, S.R., et al. Design and synthesis of chromophores with enhanced electro-optic
activities in both bulk and plasmonic–organic hybrid devices. \textit{Materials Horizons} \textbf{9}(1), 261–270 (2022).

[18] Burla, M., Hoessbacher, C., Heni, W., Haffner, C., Fedoryshyn, Y., Werner, D., Watanabe, T., Massler, H., Elder, D.L., Dalton, L.R., et al. 500 ghz plasmonic mach-zehnder modulator enabling sub-thz microwave photonics. \textit{APL Photonics} \textbf{4}(5), 056106 (2019).

[19] Alloatti, L., Palmer, R., Diebold, S., Pahl, K.P., Chen, B., Dinu, R., Fournier, M., Fedeli, J.-M., Zwick, T., Freude, W., et al. 100 ghz silicon–organic hybrid modulator. \textit{Light: Science $\&$ Applications} \textbf{3}(5), 173–173 (2014).

[20] Lu, G.-W., Hong, J., Qiu, F., Spring, A.M., Kashino, T., Oshima, J., Ozawa, M.-a., Nawata, H., Yokoyama, S. High-temperature-resistant silicon-polymer hybrid modulator operating at up to 200 gbits-1 for energy-efficient datacentres and harsh-environment applications. \textit{Nature communications} \textbf{11}(1), 1–9 (2020).

[21] Schwarzenberger, A., Kuzmin, A., Eschenbaum, C., Füllner, C., Mertens, A., Johnson, L., Elder, D., Hammond, S., Dalton, L., Randel, S., et al. Cryogenic operation of a silicon-organic hybrid (soh) modulator at 50 gbit/s and 4 k ambient temperature. In: \textit{IEEE 2022 European Conference on Optical Communication (ECOC)}, pp. 1–6 (2022). 

[22] Habegger, P., Horst, Y., Koepfli, S.M., Kohli, M., De Leo, E., Bisang, D., Destraz, M., Tedaldi, V., Meier, N., Del Medico, N., et al. Plasmonic 100-ghz electro-optic modulators for cryogenic applications.
In: \textit{European Conference and Exhibition on Optical Communication, Optica Publishing Group}, pp. 1–1 (2022).
 
[23] Hammond, S.R., OMalleya, K.M., Xub, H., Elder, D.L., Lewis, E.J. Organic electro-optic materials
combining extraordinary nonlinearity with exceptional stability to enable commercial applications. In: \textit{SPIE Photonics West} \textbf{11998}, pp. 56–66, (2022). 

[24] Teng, C., Mortazavi, M., Boudoughian, G. Origin of the poling-induced optical loss in a nonlinear optical polymeric waveguide. \textit{Applied Physics Letters} \textbf{66}(6), 667–669 (1995).

[25] Taghavi, I., Dehghannasiri, R., Fan, T., Tofini, A., Moradinejad, H., Efterkhar, A.A., Shekhar, S., Chrostowski, L., Jaeger, N.A., Adibi, A. Enhanced poling and infiltration for highly efficient electro-optic polymer-based mach-zehnder modulators. \textit{Optics Express} \textbf{30}(15), 27841–27857 (2022).

[26] Schulz, K.M., Prorok, S., Jalas, D., Marder, S.R., Luo, J., Jen, A.K.-Y., Zierold, R., Nielsch, K., Eich,
M. Mechanism that governs the electro-optic response of second-order nonlinear polymers on silicon
substrates. \textit{Optical Materials Express} \textbf{5}(8), 1653–1660 (2015).

[27] Wang, C.-T., Li, Y.-C., Yu, J.-H., Wang, C.Y., Tseng, C.-W., Jau, H.-C., Chen, Y.-J., Lin, T.-H. Electrically tunable high q-factor micro-ring resonator based on blue phase liquid crystal cladding. \textit{Optics express} \textbf{22}(15), 17776–17781 (2014).

[28] Zhang, Z., You, Z., Chu, D. Fundamentals of phase-only liquid crystal on silicon (lcos) devices. \textit{Light: Science $\&$ Applications} \textbf{3}(10), 213–213 (2014).

[29] Li, J., Chu, D. Liquid crystal-based enclosed coplanar waveguide phase shifter for 54–66 ghz applications. \textit{Crystals} \textbf{9}(12), 650 (2019).

[30] Ptasinski, J., Kim, S.W., Pang, L., Khoo, I.-C., Fainman, Y. Optical tuning of silicon photonic structures with nematic liquid crystal claddings. \textit{Optics Letters} \textbf{38}(12), 2008–2010 (2013).

[31] Lavrentovich, O.D. Ferroelectric nematic liquid crystal, a century in waiting. \textit{Proceedings of the National Academy of Sciences} \textbf{117}(26), 14629–14631 (2020).

[32] Chen, X., Korblova, E., Dong, D., Wei, X., Shao, R., Radzihovsky, L., Glaser, M.A., Maclennan, J.E., Bedrov, D., Walba, D.M., et al. First-principles experimental demonstration of ferroelectricity in a thermotropic nematic liquid crystal: Polar domains and striking electro-optics. \textit{Proceedings of the National Academy of Sciences} \textbf{117}(25), 14021–14031 (2020).

[33] Kumari, P., Basnet, B., Wang, H., Lavrentovich, O.D. Ferroelectric nematic liquids with conics. \textit{Nature Communications} \textbf{14}(1), 748 (2023).

[34] Folcia, C.L., Ortega, J., Vidal, R., Sierra, T., Etxebarria, J.: The ferroelectric nematic phase. an optimum liquid crystal candidate for nonlinear optics. \textit{Liquid Crystals} \textbf{49}(6), 899–906 (2022).

[35] Xia, R., Zhao, X., Li, J., Lei, H., Song, Y., Peng, W., Zhang, X., Aya, S., Huang, M. Achieving enhanced second-harmonic generation in ferroelectric nematics by doping d–π–a chromophores. \textit{Journal of Materials Chemistry C} \textbf{11}(32), 10905–10910 (2023).

[36] Chen, X., Korblova, E., Glaser, M.A., Maclennan, J.E., Walba, D.M., Clark, N.A. Polar in-plane surface orientation of a ferroelectric nematic liquid crystal: Polar monodomains and twisted state electro-optics. \textit{Proceedings of the National Academy of Sciences} \textbf{118}(22), p.e2104092118 (2021).

[37] Thomaschewski, M., Zenin, V.A., Wolff, C., Bozhevolnyi, S.I. Plasmonic monolithic lithium niobate
directional coupler switches. \textit{Nature communications} \textbf{11}(1), 748 (2020).

[38] Witmer, J.D., McKenna, T.P., Arrangoiz-Arriola, P., Van Laer, R., Wollack, E.A., Lin, F., Jen, A.K., Luo, J., Safavi-Naeini, A.H. A silicon-organic hybrid platform for quantum microwave-to-optical transduction. \textit{Quantum Science and Technology} \textbf{5}(3), 034004 (2020).

[39] Shekhar, S., Bogaerts, W., Chrostowski, L., Bowers, J.E., Hochberg, M., Soref, R., Shastri, B.J.
Roadmapping the next generation of silicon photonics. \textit{Nature Communications} \textbf{15}(1), 751 (2024).

[40] Singh, J., Morison, H., Guo, Z., Marquez, B.A., Esmaeeli, O., Prucnal, P.R., Chrostowski, L., Shekhar,
S., Shastri, B.J. Neuromorphic photonic circuit modeling in verilog-a. \textit{APL Photonics} \textbf{7}(4) (2022).

[41] Kim, J.-H., Aghaeimeibodi, S., Carolan, J., Englund, D., Waks, E. Hybrid integration methods for on-chip quantum photonics. \textit{Optica} \textbf{7}(4), 291–308 (2020).

[42] Ding, R., Baehr-Jones, T., Liu, Y., Bojko, R., Witzens, J., Huang, S., Luo, J., Benight, S., Sullivan, P., Fedeli, J., et al. Demonstration of a low v π l modulator with ghz bandwidth based on electro-optic
polymer-clad silicon slot waveguides. \textit{Optics Express} \textbf{18}(15), 15618–15623 (2010).

[43] Melikyan, A., Alloatti, L., Muslija, A., Hillerkuss, D., Schindler, P.C., Li, J., Palmer, R., Korn, D.,
Muehlbrandt, S., Van Thourhout, D., et al. High-speed plasmonic phase modulators. \textit{Nature Photonics} \textbf{138}(3), 229–233 (2014).

[44] Pan, Z., Xu, X., Chung, C.-J., Dalir, H., Yan, H., Chen, K., Wang, Y., Jia, B., Chen, R.T. High-speed
modulator based on electro-optic polymer infiltrated subwavelength grating waveguide ring resonator. \textit{Laser $\&$ Photonics Reviews} \textbf{12}(6), 1700300 (2018).

[45] Inoue, S.-i., Otomo, A. Electro-optic polymer/silicon hybrid slow light modulator based on one-
dimensional photonic crystal waveguides. \textit{Applied Physics Letters} \textbf{103}(17), 171101 (2013).

[46] Ummethala, S., Kemal, J.N., Alam, A.S., Lauermann, M., Kuzmin, A., Kutuvantavida, Y., Nandam, S.H., Hahn, L., Elder, D.L., Dalton, L.R., et al. Hybrid electro-optic modulator combining silicon photonic slot waveguides with high-k radio-frequency slotlines. \textit{Optica} \textbf{8}(4), 511–519 (2021).

[47] Heni, W., Haffner, C., Elder, D.L., Tillack, A.F., Fedoryshyn, Y., Cottier, R., Salamin, Y., Hoessbacher, C., Koch, U., Cheng, B., Robinson, B., Dalton, L.R., Leuthold, J. Nonlinearities of organic electro-optic materials in nanoscale slots and implications for the optimum modulator design. \textit{Optics express} \textbf{25}(3), 2627–2653 (2017). 

[48] Wang, G., Baehr-Jones, T., Hochberg, M., Scherer, A. Design and fabrication of segmented, slotted
waveguides for electro-optic modulation. \textit{Applied Physics Letters} \textbf{91}(14), 143109 (2007).

[49] Hochberg, M., Baehr-Jones, T., Walker, C., Witzens, J., Gunn, L.C., Scherer, A. Segmented waveguides in thin silicon-on-insulator. \textit{JOSA B} \textbf{22}(7), 1493–1497 (2005).

[50] Gould, M., Baehr-Jones, T., Ding, R., Huang, S., Luo, J., Jen, A.K.-Y., Fedeli, J.-M., Fournier, M.,
Hochberg, M. Silicon-polymer hybrid slot waveguide ring-resonator modulator. \textit{Optics express} \textbf{19}(5), 3952–3961 (2011).

[51] Xu, H., Liu, F., Elder, D.L., Johnson, L.E., Coene, Y., Clays, K., Robinson, B.H., Dalton, L.R. Ultrahigh
electro-optic coefficients, high index of refraction, and long-term stability from diels–alder cross-linkable binary molecular glasses. \textit{Chemistry of Materials} \textbf{32}(4), 1408–1421 (2020).

[52] Sebastián, N., Čopič, M., Mertelj, A. Ferroelectric nematic liquid-crystalline phases. \textit{Physical Review E} \textbf{106}(2), 021001 (2022).

[53] Witt, D., Young, J., Chrostowski, L. Reinforcement learning for photonic component design. \textit{APL Photonics} \textbf{8}(10) (2023).

[54] Darcie, A., Mitchell, M., Awan, K., Abdolahi, M., Hammood, M., Pfenning, A., Yan, X., Afifi, A., Witt, D., Lin, B., et al. Siepicfab: the canadian silicon photonics rapid-prototyping foundry for integrated optics and quantum computing. In: \textit{SPIE Silicon Photonics XVI}, vol. \textbf{11691}, pp. 31–50 (2021). 

[55] Wang, Y., Yun, H., Lu, Z., Bojko, R., Shi, W., Wang, X., Flueckiger, J., Zhang, F., Caverley, M., Jaeger, N.A., et al. Apodized focusing fully etched subwavelength grating couplers. \textit{IEEE Photonics Journal} \textbf{7}(3), 1–10 (2015).

[56] Yamaguchi, Y., Dat, P.T., Takano, S., Motoya, M., Hirata, S., Kataoka, Y., Ichikawa, J., Oikawa, S., Shimizu, R., Yamamoto, N., et al. Traveling-wave mach–zehnder modulator integrated with electro-optic frequency-domain equalizer for broadband modulation. \textit{Journal of Lightwave Technology} \textbf{41}(12), pp.3883-3891 (2023).

[57] Luan, E., Yu, S., Salmani, M., Nezami, M.S., Shastri, B.J., Chrostowski, L., Eshaghi, A. Towards a
high-density photonic tensor core enabled by intensity-modulated microrings and photonic wire bonding. \textit{Scientific Reports} \textbf{13}(1), 1260 (2023).

[58] Mitchell, M., Lin, B., Taghavi, I., Yu, S., Witt, D., Awan, K., Gou, S., Young, J., Chrostowski, L. Photonic wire bonding for silicon photonics iii-v laser integration. In: \textit{2021 IEEE 17th International Conference on Group IV Photonics (GFP)}, pp. 1–2 (2021). 

[59] Wang, T., Lin, B., Mitchell, M., Taghavi, I., Chrostowski, L., Jaeger, N.A. Semiconductor optical amplifier (soa) integrated on a silicon photonic chip using photonic wire bonds (pwbs). In: \textit{SPIE Integrated Optics: Devices, Materials, and Technologies XXVIII}, vol. \textbf{12889}, pp. 131–137 (2024). 

[60] Palmer, R., Koeber, S., Elder, D.L., Woessner, M., Heni, W., Korn, D., Lauermann, M., Bogaerts, W., Dalton, L., Freude, W., et al. High-speed, low drive-voltage silicon-organic hybrid modulator based on a binary-chromophore electro-optic material. \textit{Journal of Lightwave Technology} \textbf{32}(16), 2726–2734 (2014).

[61] Koeber, S., Palmer, R., Lauermann, M., Heni, W., Elder, D.L., Korn, D., Woessner, M., Alloatti, L., Koenig, S., Schindler, P.C., et al. Femtojoule electro-optic modulation using a silicon–organic hybrid device. \textit{Light: Science $\&$ Applications} \textbf{4}(2), 255–255 (2015).

[62] Jin, W., Johnston, P.V., Elder, D.L., Tillack, A.F., Olbricht, B.C., Song, J., Reid, P.J., Xu, R., Robinson, B.H., Dalton, L.R. Benzocyclobutene barrier layer for suppressing conductance in nonlinear optical devices during electric field poling. \textit{Applied Physics Letters} \textbf{104}(24), 94–1 (2014).

[63] Elder, D.L., Haffner, C., Heni, W., Fedoryshyn, Y., Garrett, K.E., Johnson, L.E., Campbell, R.A., Avila,
J.D., Robinson, B.H., Leuthold, J., et al. Effect of rigid bridge-protection units, quadrupolar interactions, and blending in organic electro-optic chromophores. \textit{Chemistry of Materials} \textbf{29}(15), 6457–6471 (2017).

[64] Taghavi, I., Dehghannasiri, R., Fan, T., Moradinejad, H., Taghinejad, H., Hosseinnia, A.H., Eftekhar, A.A., Adibi, A. Enhanced polling and infiltration of highly-linear mach-zehnder modulators on si/sin-organic hybrid platform. In: \textit{CLEO: Science and Innovations, Optical Society of America}, pp. 1–1 (2018). 

[65] Brosi, J.-M., Koos, C., Andreani, L.C., Waldow, M., Leuthold, J., Freude, W. High-speed low-voltage
electro-optic modulator with a polymer-infiltrated silicon photonic crystal waveguide. \textit{Optics Express} \textbf{16}(6), 4177–4191 (2008).

[66] Kamada, S., Ueda, R., Yamada, C., Tanaka, K., Yamada, T., Otomo, A. Superiorly low half-wave voltage electro-optic polymer modulator for visible photonics. \textit{Optics Express} \textbf{30}(11), 19771–19780 (2022).

[67] Onural, D., Wang, I., Chiang, L.-Y., Raja, S., Zhang, X., Singh, M., Li, D., Dao, H., Pajk, S., Sickler, J.W., et al.: Hybrid integration of silicon slot photonics with ferroelectric nematic liquid crystal for poling-free pockels-effect modulation. \textit{In: 2024 Conference on Lasers and Electro-Optics (CLEO), IEEE}, pp. 1–2 (2024).
 
[68] Park, J.W., You, J.-B., Kim, I.G., Kim, G. High-modulation efficiency silicon mach-zehnder optical modulator based on carrier depletion in a pn diode. \textit{Optics express} \textbf{17}(18), 15520–15524 (2009).

[69] Qiu, F., Sato, H., Spring, A.M., Maeda, D., Ozawa, M.-a., Odoi, K., Aoki, I., Otomo, A., Yokoyama, S. Ultra-thin silicon/electro-optic polymer hybrid waveguide modulators. \textit{Applied Physics Letters} \textbf{107}(12),92–1 (2015).

[70] Zhang, Xingyu and Lee, Beomsuk and Lin, Che-yun and Wang, Alan X and Hosseini, Amir and Chen, Ray T Highly linear broadband optical modulator based on electro-optic polymer. \textit{IEEE Photonics Journal} \textbf{4}(6),2214-2228(2012).

\begin{spacing}{1.2}
\large

\section*{Acknowledgements}
The Natural Sciences and Engineering Research Council of Canada (NSERC), the SiEPICfab consortium, the B.C. Knowledge Development Fund (BCKDF), the Canada Foundation for Innovation (CFI), MITACS funding with Dream Photonics Inc., and the Schmidt Sciences funded this research.

\section*{Author contribution}

This work was conceptualized by I.T., L.C., and S.S. C.P., J.S., J.Y., and N.J. contributed materials and technical information. At the same time, M.H. helped with design and S.J., K.M.A., D.W., and I.T. built the organic Si phase shifters described in this paper. I.T. conducted measurements while N.J. and O.E. helped with the modulator measurement setup. The manuscript was written by I.T., L.C., and S.S., with assistance from all authors.  

\section*{Competing interests}
The authors declare no competing interests.

\newpage

\begin{table}[h!]
\centering
\begin{adjustbox}{max width=\textwidth}
\begin{tabular}{*{5}{c}}
\toprule
\Centerstack{\textbf{} } &
\Centerstack{EO polymer} &
\Centerstack{Organic NLO$^{1}$} &
\Centerstack{PN-LC$^{2}$} &
\Centerstack{FN-LC$^{3}$} 
\\\hline
Non-zero second-order susceptibility, $\upchi^{(2)}$& \textcolor{lightgray}{$\CIRCLE$} &\textcolor{lightgray}{$\CIRCLE$} & &$\CIRCLE$\\
Non-centrosymmetric crystal lattice&\textcolor{lightgray}{$\CIRCLE$} & \textcolor{lightgray}{$\CIRCLE$} &     &$\CIRCLE$\\
Spontaneous molecular organization & & \textcolor{lightgray}{$\CIRCLE$} &\textcolor{lightgray}{$\CIRCLE$} &$\CIRCLE$\\
Alignments using interface layer or small $\textbf{E}_\text{x}$ &  &   &\textcolor{lightgray}{$\CIRCLE$}  &$\CIRCLE$\\
Reorientation of slow/fast axes in response to external $\textbf{E}_\text{x}$  &  &   & \textcolor{lightgray}{$\CIRCLE$} &$\CIRCLE$\\
Large spontaneous polarizations and second order nonlinearities &  &   & &$\CIRCLE$\\
\bottomrule
\end{tabular}
\end{adjustbox}
\caption{OEO material comparison}
\begin{tablenotes}
\item[1] Nonlinear optic
\item[2] Paraelectric nematic liquid crystal
\item[3] Ferroelectric nematic liquid crystal
\end{tablenotes}
\label{OEO}
\end{table}

\begin{table}[h!]
\centering
  
\begin{adjustbox}{max width=\textwidth}
\begin{tabular}{*{9}{c}}
\toprule
\Centerstack{Topology \\  } &
\Centerstack{Phase shift \\ material} &
\Centerstack{Size$^{2}$ \\ (mm)} &
\Centerstack{V$_{\pi}$ \\ (V)} &
\Centerstack{V$_\pi L\upalpha$ \\ (V.dB)} &
\Centerstack{$ER_\text{opt}$ \\ (dB)} &
\Centerstack{$\partial n / \partial v$\\ (V$^{-1}$)} &
\Centerstack{$\partial \upkappa / \partial v$\\ (V$^{-1}$)} &
\Centerstack{Ref. \\}
\\\hline
            MZM         &LNOI    &3   &7.4  &2.18 &40  & 6.98$\times10^{-5}$ &-                  & [9]\\  
            MZM$^{1}$   &ITO     &0.03&16   &80   &-   & 1.47$\times10^{-3}$ &1.64$\times10^{-3}$& [8]\\
            MZM         &BTO     &2   &1.15 &1.3  &-  & 3.37$\times10^{-4}$ &1.37$\times10^{-5}$ &[12]\\
            MZM         &doped Si&2   &9    &79.2 &12.5& 3.89$\times10^{-4}$ &$2.54\times10^{-5}$& [68]\\
            nonslot-MZM    &EO polymer &13  &4.6  &23.92&-   & 2.59$\times10^{-5}$ &-& [69]\\
            nonslot-MZM    &EO polymer &8   &1.8  &3.17 &-   & 1.07$\times10^{-4}$ &-& [20]\\
            FLS-MZM     &FN-LC  &0.5 & 0.5$^{*}$ (51.4$^{**}$) & {1.05}$^{*}$ (107.94$^{**}$) &26  &4.04$\times10^{-3}$  &4.54$\times10^{-5}$ & this work\\


\bottomrule
\end{tabular}
\end{adjustbox}
\caption{Modulators comparison}
\begin{tablenotes}
\item[1] Electro-absoprtion modulator
\item[2] Arm length of a Mach-Zehnder Modulator (MZM)
\item[*] DC metrics
\item[**] AC metrics
\end{tablenotes}
\label{FOMtable}
\end{table}

\end{spacing}
\end{document}